\documentclass[10pt,twocolumn,amsmath,amssymb,aps,pra,showpacs]{revtex4-2}
\usepackage{mathrsfs}
\usepackage{graphicx}
\usepackage{bm}
\usepackage{amssymb}
\usepackage{amsmath}
\usepackage{paralist}
\usepackage{amsmath,amssymb,mathrsfs}

\makeatletter

\renewcommand{\maketag@@@}[1]{\hbox{\m@th\normalsize\normalfont#1}}%

\makeatother
\usepackage{color}

\begin{document}

\title{One-body dynamical correlation function of the Lieb-Liniger model at finite temperature}

\author{Song Cheng$^{1,2,3}$}

\author{Yang-Yang Chen$^{1,4}$}
\email[]{chenyy@nwu.edu.cn}

\author{Xi-Wen Guan$^{5,6,7}$}

\author{Wen-Li Yang$^{1,4,7}$}

\author{Hai-Qing Lin$^{3,8}$}
\email[]{haiqing0@csrc.ac.cn}

\affiliation{$^1$Institute of Modern Physics, Northwest University, Xi'an 710069, China}

\affiliation{$^2$ Department of Physics, The University of Hong Kong, Pokfulam Road, Hong Kong, China}

\affiliation{$^3$Beijing Computational Science Research Center, Beijing 100193, China }

\affiliation{$^4$Shaanxi Key Laboratory for Theoretical Physics Frontiers, Xi’an 710069, China}

\affiliation{$^5$5Innovation Academy for Precision Measurement Science and Technology, Chinese Academy of Sciences, Wuhan 430071, China}

\affiliation{$^6$ Department of Fundamental and Theoretical Physics, Research School of Physics, Australian National University, Canberra ACT 0200, Australia}

\affiliation{$^7$Peng Huanwu Center for Fundamental Theory, Xi'an 710069, China}

\affiliation{$^8$School of Physics, Zhejiang University, Hangzhou 310058, China}



\begin{abstract}
The dynamical correlated properties of one-dimensional (1D) Bose gases provide profound understanding of novel physics emergent from collective excitations, for instance, the breakdown of off-diagonal long-range order, and the establishment of Tomonaga-Luttinger liquid theory. However, due to the nonperturbative nature of 1D many-body systems, the exact evaluation of correlation functions is notoriously difficult. Here, by means of a form factor approach based on an algebraic Bethe ansatz and numerics, we present a thorough study on the one-body dynamical correlation function (1BDCF) of the Lieb-Liniger model at finite temperature.
The influence of thermal fluctuation and interaction on the behavior of 1BDCF has been demonstrated and analyzed from various perspectives, including the spectral distribution, the line shape at fixed momentum, and the corresponding static correlations.
\end{abstract}

\maketitle

\noindent

In the past two decades, a growing interest in low-dimensional quantum many-body systems has stimulated tremendous success in both theoretical and experimental  developments of quantum physics  \cite{Cazalilla,Imambekov2012,Guan2013,Guan2022,Cazalilla2004}.
In contrast to the emergent quasiparticle in Landau-Fermi liquid theory for condensed-matter problems in higher dimensions \cite{Landau:1957,Anderson:1984},
the bosonic collective excitations \cite{Haldane:1981,Giamarchi} take over the main role of low-energy physics for the one-dimensional (1D) strongly correlated systems.
Consequently, significantly different phenomena away from the
quasiparticle picture occur at low temperature in 1D interacting systems, such as the absence of true Bose-Einstein condensate (BEC) in homogeneous Bose gases \cite{Giorgini:2008}, fractional spin and holon excitations \cite{Ess05}, and spin-charge separation \cite{Hilker:2017,Vijayan:2020,Senaratne:2022}.
Moreover, the presence of quantum integrability in 1D results in the breakdown of thermalization for quantum gases \cite{Kinoshita:2006,Langen:2015,Wilson2020}.
The key to understanding such unique physics lies in the correlation functions of either ground state or finite temperatures, yet the nonperturbative nature brings it with a big challenge on their exact evaluation \cite{Giamarchi}.
In this Letter, making use of quantum integrable theory, we thoroughly study the finite-temperature one-body dynamical correlation function (1BDCF) of the Lieb-Liniger model with an arbitrary interaction.

The Lieb-Liniger model describes $N$ spinless bosons confined  in a line of length $L$ via contact interaction \cite{Lieb} and its Hamiltonian reads
\begin{equation}\label{hamiltonian}
H = - \sum_{i=1}^{N} \frac{\partial^2}{\partial x_i^2} + 2c \sum_{i>j}^{N} \delta \left( x_i - x_j \right),
\end{equation}
where $c>0$ ($c<0$) specifies the interaction strength of repulsion (attraction) and hereafter we only consider the repulsive case.
The physics of this model is governed by two dimensionless parameters simultaneously, i.e.,  interaction strength $\gamma = c/n$  and
temperature $\tau=T/T_d$. Here we denote the  particle density $n=N/L$ and degenerate temperature $T_d=\hbar^2n^2/(2mk_B)$, respectively.
By tuning these parameters the system would vary across several different regimes: quasicondensate, Tonks-Girardeau (TG) gas, degenerate regime, decoherence regime, quantum criticality, etc. \cite{Kheruntsyan,Guan:2011}.
In the TG limit $\gamma \rightarrow \infty$,  the one-body density matrix was derived at zero and finite temperatures by the trick of the Bose-Fermi map \cite{Girardeau,Lenard,Tracy}, as well as in the presence of trap potential \cite{Colcelli2018a,Colcelli2018b,Colcelli2020,Pezer,Minguzzi}.
Away from this limit, on the basis of pseudopotential Hamiltonian and a combination of generalized Hartree-Fock approximation and random phase approximation, the dynamic structure factor (DSF) in the strongly correlated region was obtained up to a correction of $1/\gamma$ order \cite{Cherny}.
While with the application of Hellmann-Feynman theorem, the pair-correlation function was calculated  \cite{Gora_a,Kheruntsyan}, a general expression for higher-order static correlation function in a strongly correlated regime was found through the asymptotic Bethe ansatz solution \cite{Gora_b,Nandani}.
Besides, a variety of static correlated properties were studied through the quantum Monte Carlo method \cite{Xu2015,Rosi2023}, while the edge singularity of dynamical correlation functions was predicted by the nonlinear Tomonaga-Luttinger liquid (TLL) theory \cite{Imambekov2008,Imambekov2012}.

Being the simplest quantum integrable system \cite{Lieb}, the Lieb-Liniger model has been acting as a major subject of textbook materials for quantum many-body physics \cite{Giamarchi,Franchinni,QISM}.
Nevertheless, the study of correlation functions  is still notoriously difficult  due to complexity of many-body eigenstates \cite{QISM}.
Towards this end,  much effort has  been devoted to finding the determinant representations of various correlators through the algebraic Bethe ansatz (ABA) technique \cite{Korepin1982,Korepin1984,Slavnov,Slavnov1997}.

These tedious formulas represent a big challenge for a full access to the correlated properties and thus intrigue the numerical methods that count in useful eigenstates and calculating the determinant expressions of form factors, such as ABACUS \cite{Caux2006,Caux2007,Panfil,Caux2009,Caux2023} and our recent achievements \cite{SC,YYC2023}.
This technique provides a theoretical benchmark for the cold-atom experiments on correlated properties in 1D systems \cite{Exp}.
On the other hand, by restricting the number of pairs of particle-hole excitation, the DSF and spectral function were obtained for the low-density region by utilizing the thermodynamic form factor \cite{Granet}.
The quantum integrability technique enables evaluation of the nonuniversal prefactors present in bosonization theory as well \cite{Shashi}.

Here we present a thorough study on 1BDCF of the Lieb-Liniger model at zero and finite temperatures with arbitrary interaction strengths, by utilizing a form factor approach on basis of ABA and numerics.
The spectral distributions, the line shape of 1BDCF, the momentum distribution, and the one-body density matrix are obtained to uncover the dynamical correlated properties of the Lieb-Liniger model at finite temperatures.

\begin{figure}[ht]
\includegraphics[width=0.5\textwidth]{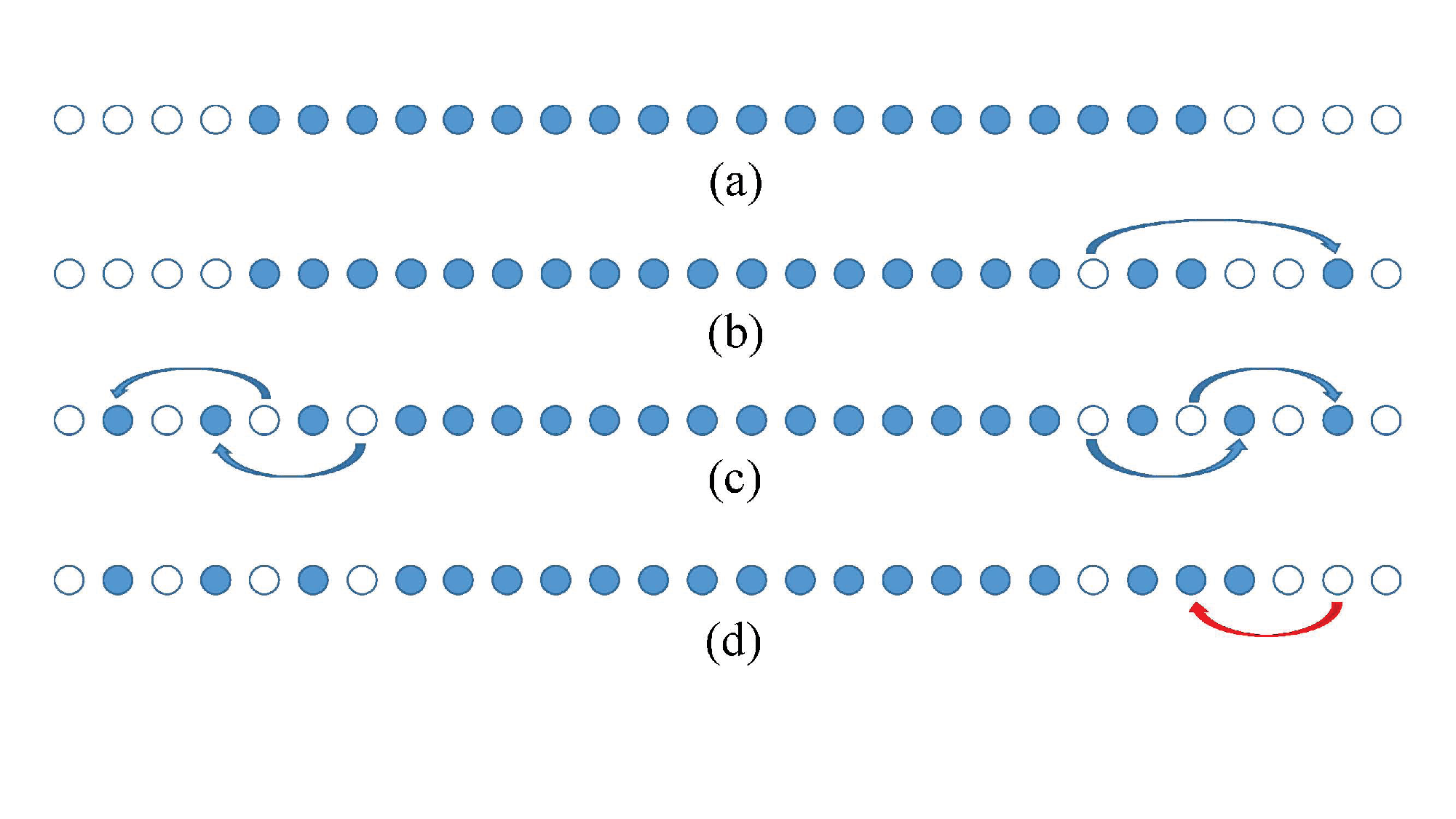}
\caption{Illustration of quantum numbers and excitation. Panel (a) depicts the configuration of QNs for the  ground state, an analogy of Fermi sea for noninteracting spinless fermions in 1D. For a $N$-particle system, the QNs in the ground state  consecutively range from $-(N-1)/2$ to $(N-1)/2$.
Panel (b) shows an example of conventional $1$-pair of particle-hole (p-h) excitation.  Panel (c) demonstrates how a reference state is achieved by means of  multi-pairs of p-h excitation, where the configuration of QNs is obviously symmetric. Panel (d) shows an example of "relative excitation" over the original reference state displayed in (c), giving birth to states responsible for the secondary peak in Fig.~\ref{lineshape}.}
\label{excitation}
\end{figure}

By virtue of periodical boundary conditions and Bethe wave function, the diagonalization of the Hamiltonian in Eq.~(\ref{hamiltonian}) is equivalent  to solving a set of transcendental  equations, i.e., Bethe ansatz equations (BAEs) \cite{Lieb},
\begin{equation}\label{BAE}
\lambda_j + \frac{1}{L} \sum_{k=1}^{N} \theta \left( \lambda_j - \lambda_k \right) = \frac{2\pi}{L} I_j, \quad j = 1, \dots, N
\end{equation}
where $\theta(x)=2\arctan(x/c)$, and $\lambda_j$ and $I_j$ are the pseudomomentum and  corresponding quantum number (QN), respectively.
A set of $\{ I_j \}$ uniquely determines a quantum state (represented by a set of $\{\lambda_j\}$) and vice versa. 
Those QNs take an integer (half-integer) if $N$ is odd (even).
The total momentum and energy of the system are expressed in terms of pseudomomenta,
$P_{\{\lambda\}}=\sum_{j=1}^{N} \lambda_j$ and $E_{\{\lambda\}}=\sum_{j=1}^{N} \lambda_j^2$.
It is convenient to illustrate the eigenstates of the system by making use of the configuration for QNs. The ground state is depicted by a Fermi sea--like configuration, over which pairs of particle-hole (p-h) excitation simply generate excited states; see Figs.~\ref{excitation}(a) and \ref{excitation}(b). This configuration of QNs in the thermodynamic limit becomes a continuous function of pseudomomentum.

\begin{figure}[!htbp]
\includegraphics[width=0.5\textwidth]{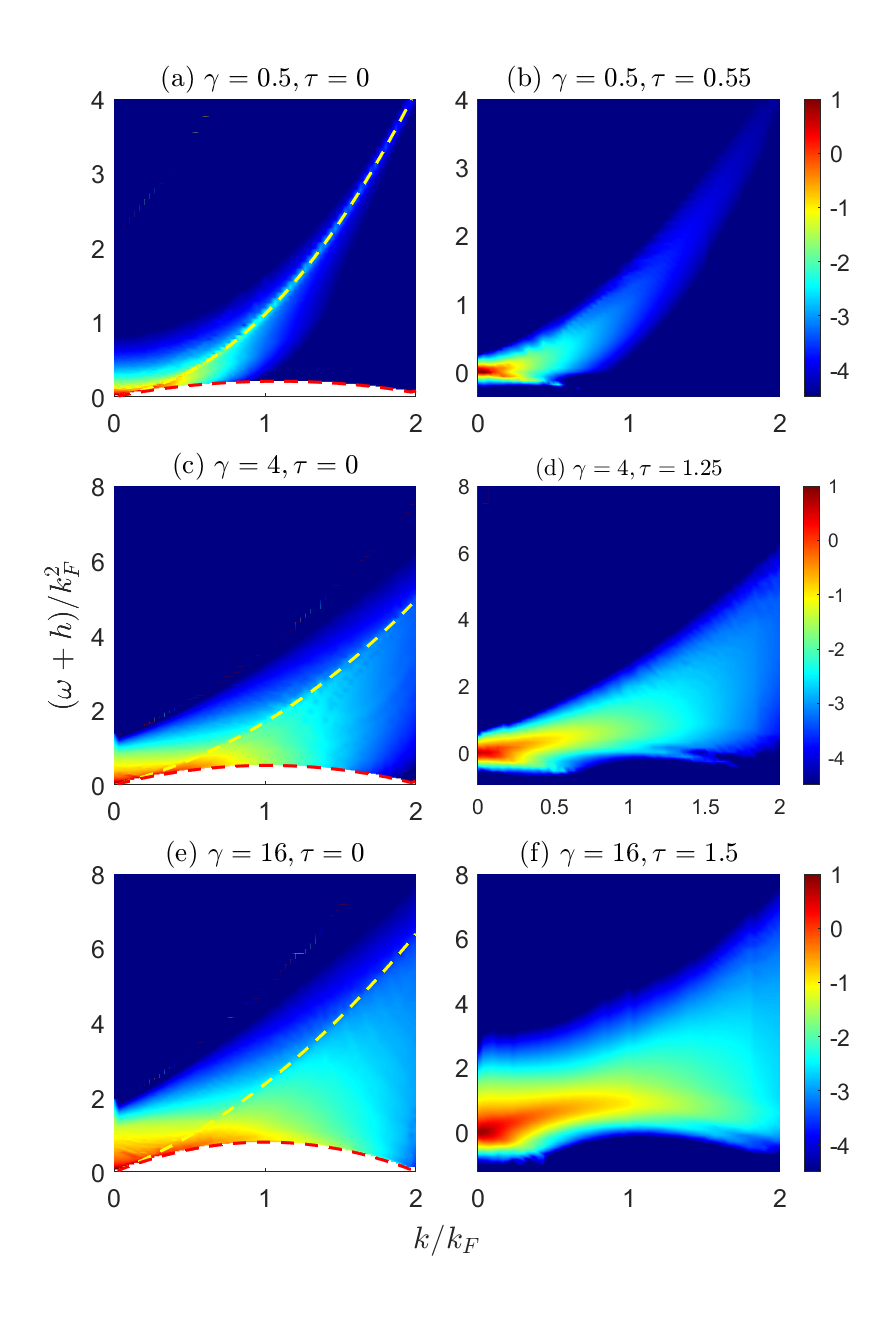}
\caption{Logarithmic 1BDCF at $k$ - $\omega$ plane with different temperature $\tau$ and interaction $\gamma$ of system size $N=L=60$. $h$ is the chemical potential, 
while momentum (energy) is measured in units of Fermi momentum (Fermi energy).
The left panel is for zero temperature and the right for finite temperature.
The yellow- and red-dashed curves respectively stand for the type-I and -II dispersions.
The sum rules [see Eq.~(\ref{sum_rule})] of (a)-(f) are 99.99\%,  99.83\%, 99.63\%, 97.63\%,  98.90\%, and 98.99\%, respectively.}
\label{color_plot}
\end{figure}

The 1BDCF of our interest is defined by
\begin{equation}\label{correlation0_def}
\resizebox{0.9\hsize}{!}{$
g_1 (k,\omega) = \int_{0}^{L} \textmd{d}x  \int_{-\infty}^{\infty} \textmd{d}t \, e^{\textmd{i}(\omega t - k x)} \langle \Psi^\dagger(x,t) \Psi(0,0) \rangle_T
$}
\end{equation}
where $\Psi(x,t)$ ($\Psi^{\dagger}(x,t)$)  is the bosonic field operator of  annihilation (creation)  and $\langle \cdots \rangle_T$ stands for a statistical expectation at finite temperature $T$.
The standard treatment for this ensemble average  needs computation  of the partition function, which however is extremely hard to conduct for a strongly correlated system. 
It turns out this difficulty  can be circumvented by means of quantum integrability, namely by the thermodynamic Bethe ansatz (TBA) equations \cite{QISM,YY,Panfil}. There it makes sense to speak of an ensemble that solves the TBA equations. These equations can be solved once the chemical potential, interaction strength, and temperature are specified. Meanwhile, a thermodynamic equilibrium state consists of a mixture of eigenstates sharing the same macroscopic description correspondingly.
Therefore, it is crucial to just find one arbitrary eigenstate to represent all of the states determined by the same TBA equations. The problem now is simplified and similar to the ground state. We will call this specified eigenstate henceforth "reference state" which comes from the generating process for excited states and we will discuss it later on.
In principle, the configuration of QNs for a reference state simultaneously depends on the interaction strength and temperature. For a given interaction strength $\gamma$, the higher temperature will excite more p-h pairs over the Fermi sea in the ground state; see the Supplemental Material \cite{note4SM} for how to obtain a reference state by solving the TBA equations.
After inserting a completeness relation of intermediate states 
$\sum_{\{\mu\}} | \{\mu\} \rangle \langle \{\mu\} |/ \| \{\mu\} \|^2=1$,  the 1BDCF is  given by
\begin{equation}\label{spectral_form}
\resizebox{0.9\hsize}{!}{$
g_1 \left( k,\omega \right) = 2\pi L \sum_{\{\mu\} } \frac{ \|  \langle \{\mu\} | \Psi(0,0)  |  \{\lambda\}_T  \rangle \|^2 }{\|  \{\lambda\}_T \|^2  \| \{\mu\} \|^2}  \,
\delta_{k,P_{\mu,\lambda}}  \delta\left(\omega-E_{\mu,\lambda} \right)
$}
\end{equation}
where eigenstate $| \{ \lambda \}_T \rangle$ is the reference state obtained from solving the TBA equation.
In the above equation, we denote  $O_{ \mu,\lambda} \equiv O_{\{\mu\}}- O_{\{\lambda\}}$ with $O=P$ and $E$.
$\delta_{m,n}$ and  $\delta(x)$ are the Kronecker- and Dirac-delta functions, respectively.
$\langle \{ \mu \} | \Psi(0,0)  | \{ \lambda\}_T \rangle $ is called \textit{form factor} and $\| \{ \nu \} \|^2$ is the norm square of the  eigenstate, both of which can be numerically  calculated through the determinants with entries represented by pseudomomenta  \cite{Caux2007}.

In principle, the sum is taken over the whole Hilbert space, which is however impossible in practice.
We therefore incorporate the states with significant contributions to the correlation function as much as possible until obtaining a satisfying result of the sum rule \cite{SC}.
This exploring process is equal to reproducing excited states, which of course can be fulfilled by generating 1-, 2- , \dots, and $m$-pairs of p-h excitation over the ground state.
Nevertheless, in the situation of finite temperature, this will be a waste of computation resource.
To avoid this drawback, we start the above generating process from an eigenstate derived from a TBA equation instead, explaining the meaning of \textit{reference}.
Any configuration of QNs away from this reference state will be considered as a "relative excitation," since it need not be the ground state; see Fig.~\ref{excitation}(d). More details can be found in the Supplemental Material.
We then count states by manipulating such type of "relative excitations."
This thought comes from a naive observation that the most relevant states for the form factor of a local observable should be similar looking to the reference state from the viewpoint of configuration of QNs.

For a dynamical correlation, there usually exists a sum rule that connects itself with a static quantity by integrating or summing out the frequency. 
This relation can be employed to check the validity of the dynamical correlation.
Here, for 1BDCF, the sum rule reads \cite{Caux2007}
\begin{equation}\label{sum_rule}
\sum_{k} n (k) = \frac{N}{L}
\end{equation}
where $n(k) = \int_{-\infty}^{\infty} \frac{\textmd{d}\omega}{2\pi L} g_1(k,\omega)$  is the momentum distribution. 
In our setting, the density is $n=N/L=1$ and thus the sum rule close to $1$ indicates a faithful result; see Fig.~\ref{color_plot} where the sum rules from (a) to (f) are respectively 99.99\%,  99.83\%, 99.63\%, 97.63\%,  98.90\%, and 98.99\%.

In the strongly degenerate regime ($\tau \ll 1$), it is reasonable to assume 1BDCF to be similar to its zero-temperature situation, merely differing with a temperature-dependent perturbation.
In a weakly interacting regime ($\tau \ll \gamma \ll 1 $ or $\gamma \ll \tau \ll \sqrt{\gamma}$),
Bogoliubov theory provides a way to understand some quantum correlation functions \cite{Zambelli2000}.
Away from this limit, increasing temperature gradually destroys the phase coherence and thus the system eventually  arrives at the decoherent regime \cite{Kheruntsyan}.
Here we focus on the regime  $\tau \sim \sqrt{\gamma} \sim 1$,  where the competition between thermal fluctuation and interaction recomposes the correlated properties.

The frequency-momentum resolved 1BDCF is demonstrated in Fig.~\ref{color_plot} with system size $N=L=60$. Thus the Fermi momentum $k_F = \pi n = \pi$, the momentum and energy are expressed in units of Fermi momentum and Fermi energy, respectively. $h$ is the chemical potential.
The left panel shows that at zero temperature, in the situation of weak interaction, spectral weights mostly focus on the type-I dispersion represented by the yellow dashed curves; while away from the weakly interacting region, the spectral weights mostly concentrate on the type-II dispersion represented by the red-dashed curves, and spread across the type-I dispersion.
The latter (former) dispersion corresponds to adding a particle (hole) outside (inside) of the Fermi sea \cite{note}.
In  the right panel of the finite-temperature case, immediate differences are observed with  the nonvanishing spectral distribution in the half-plane of negative energy.
This is easy to understand from the perspective of p-h excitation as being given in Fig.~\ref{excitation}.
In Fig.~\ref{excitation}(c), the reference state at finite temperature indicates a melted Fermi sea. 
Besides the normal "relative excitations" whose absolute values of particle QN increase, there exists such a case that the absolute value of particle QN decreases.
A typical example of such an "excitation" is depicted by the red arrow in Fig.~\ref{excitation}(d), which generates the states bearing nonvanishing spectral contribution while possessing lower energy than the reference state.
In the quantum degenerate regime, the reference state contains very few holes lying almost on the Fermi points, which leaves no further room for the above type of "excitations." This in consequence suppresses the spectral distribution to the negative energy plane.

\begin{figure}
\centering
\includegraphics[width=1.15\linewidth]{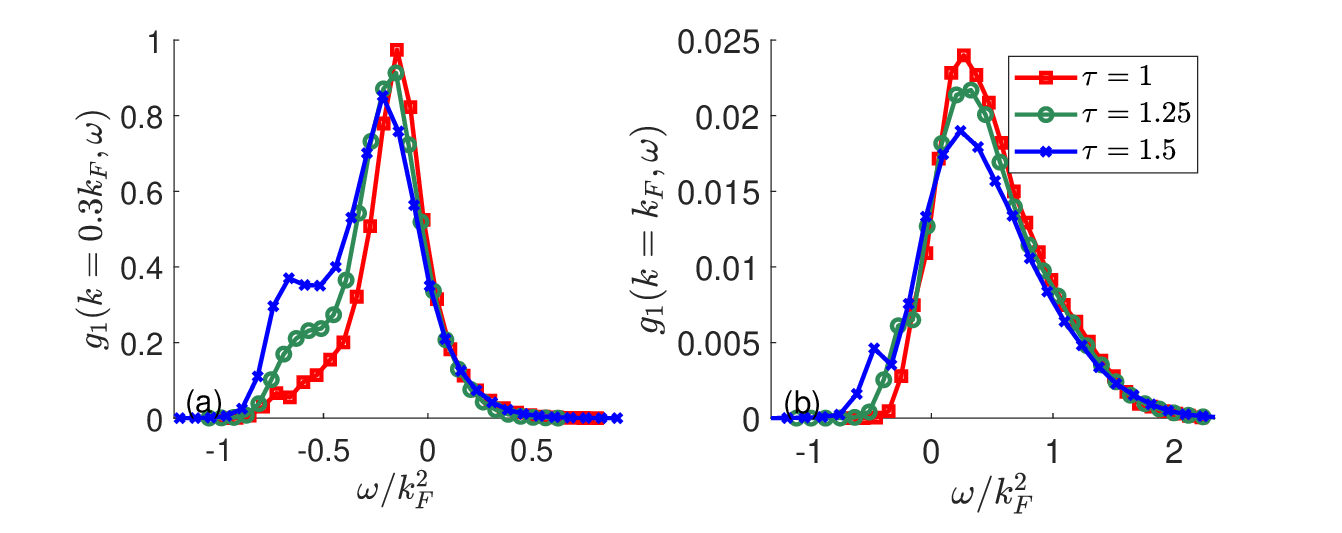}
\caption{ Panels (a) and (b) show the line shape of $g_1(k,\omega)$ vs $\omega$ with $k=0.3 k_F$ and $k=k_F$, respectively.
The system size is $N=L=60$ and the dimensionless interaction strength is $\gamma = 4$. The energy is measured in units of Fermi energy.
The red, green, and blue curves stand for the results at rescaled temperatures $\tau= 1.0$, $1.25$, and $1.5$, respectively.
The secondary peak appears due to the thermal fluctuation, whose signal becomes obvious with the increase of temperature. 
}
\label{lineshape}
\end{figure}

\begin{figure}[!htbp]
\includegraphics[width=0.48\textwidth]{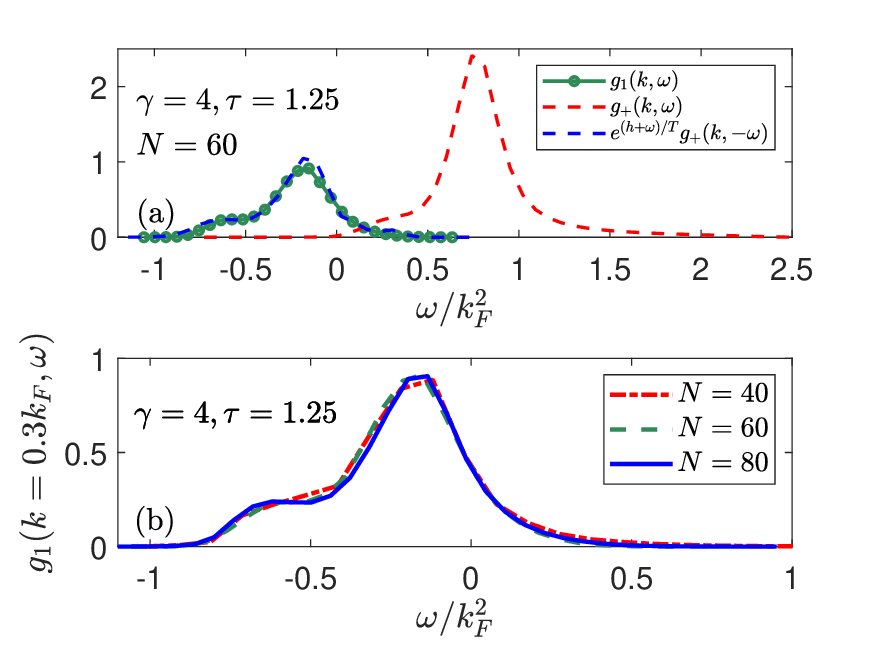}
\caption{
For $\gamma=4$, $\tau=1.25$, and $n=1$, (a) the detailed balancing relation and (b) the finite-size effect are shown.
The green and red curves of (a) respectively represent the results of $g_1(0.3k_F,\omega)$ and $g_+(0.3 k_F,\omega)$, while the blue one derives from the detailed balancing relation.
There is good agreement between the blue and green curves.
In (b) the solid blue, dashed green, and dotted red curves denote 
$g_1(0.3 k_F,\omega)$  of different system sizes $N=L=80$, $60$, and $40$, respectively.
The results from $N=L=60$ and $80$ basically agree with each other and it is safe to conduct calculation when $N=L=60$.
}
\label{finite}
\end{figure}

In Fig.~\ref{lineshape}, the line shape of 1BDCF with a given momentum is displayed  by taking the intermediate interaction strength $\gamma=4$.
At zero temperature, there is an analog of Fermi edge singularity lying on the type-II dispersion \cite{Imambekov2008,SC,Caux2007}, which is smeared for finite temperature by the thermal fluctuation.
Although the spectral weight in Fig.~\ref{lineshape} at the given momenta shrink, comparing with the counterpart in the ground state, the percentage of spectral weight occupied by the multipairs of p-h excitation become prominent. Accompanied with this process is the redistribution of spectral weight on the energetic axis, as is shown in Fig.~\ref{color_plot}.
Along with an increase of the temperature, a secondary peak emerges in the plane of negative energy gradually.  Like what is discussed before, its emergence arises from the states produced by the relative excitation, which decreases the absolute values of QN. Besides, the main peak is lowered in this process, whose loss of spectral weight is compensated partly by the secondary peak.
In fact, this double-peak phenomenon is exactly the reflection of the detailed balancing relation for the dynamic correlation functions \cite{LL}
$g_1(k,\omega)= e^{(h+\omega)/T} g_+(k,-\omega)$
where $h$ is the chemical potential and $g_+(k,\omega)=\int_{0}^{L} \textmd{d}x  \int_{-\infty}^{\infty} \textmd{d}t \, e^{\textmd{i}(\omega t - k x)} \langle \Psi(x,t) \Psi^\dagger(0,0) \rangle_T$.
A similar phenomenon is observed for DSF as well \cite{Granet,Panfil,LL}.
Figure~\ref{finite}(a) explicitly shows this relation for the line shape of 1BDCF with $\gamma=4$ and $\tau=1.25$ of system size $N=L=60$.
The green and red curves indicate $g_1(k=0.3k_F,\omega)$ and $g_+(k=0.3k_F,\omega)$, respectively, and the blue one is obtained by utilizing a detailed balancing relation, in accordance  with that of 1BDCF.
We briefly show the finite-size effect through Fig.~\ref{finite}(b), where the 1BDCF of different system sizes are compared.
It is shown that the results of $N=L=60$ and $80$ basically agree with each other, with a slight difference from that of $N=L=40$.  Hence it is convincing to make use of the system size $N=L=60$.

\begin{figure}
\centering
\includegraphics[width=0.4\textwidth]{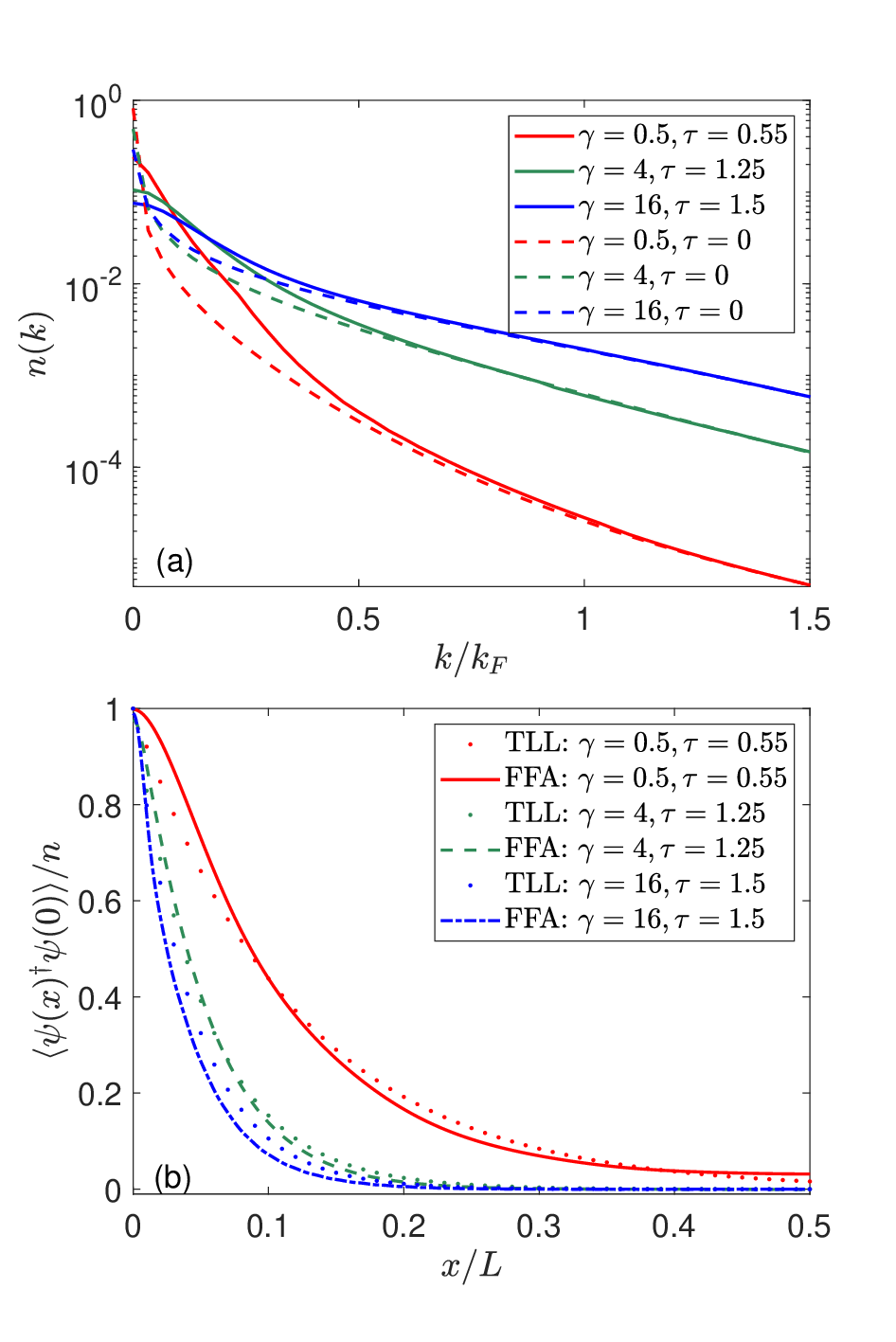}
\caption{(a) Momentum distributions at  different temperature  $\tau$ and interaction strength  $\gamma$. All data come from Fig.~\ref{color_plot}.
For the sake of a clear visibility, we take a logarithmic scale for the momentum distribution.
The dotted and solid curves stand for the situations at zero and finite temperatures,  respectively.
(b) The one-body density matrix with different  temperature $\tau$ and interaction strength  $\gamma$.
The solid and dotted lines respectively correspond to the TLL predictions and our results derived from the form factor approach (FFA).
The parameters of $(\gamma,\tau)$ for red, green, and blue curves are $(0.5,0.55), (4,1.25),$ and $(16,1.5)$.
The TLL prediction does not work in the above parameter setting since the thermal fluctuation is strong enough to break down its validity.
}
\label{static}
\end{figure}

The static correlation function corresponding to 1BDCF (i.e., the momentum distribution) is obtained by integrating out the frequency, whose Fourier transformation is the one-body density matrix.
In Fig.~\ref{static}(a) we compare the momentum distribution at zero and finite temperatures based on the data from Fig.~\ref{color_plot}.
Despite no BEC in the Lieb-Liniger model,  the weak interaction still results in a more prominent zero-momentum occupation than the larger interactions do.
A direct comparison shows obvious discrepancies between the curves of zero and finite temperatures with the same interaction strength, which is simply a result of disturbance from the thermal fluctuation on particles of small momentum.
It is also evidenced by Fig.~\ref{lineshape}, where the height of secondary peaks can be thought of as a measure to the influence of thermal fluctuation. 
Therein, at the same temperature, the curves at small momentum have higher secondary peaks, implying the stronger influence. 
Besides, it is seen that the interaction strength affects the discrepancy between results of zero and finite temperatures, i.e., the curves of larger interaction show less discrepancy.
The one-body density matrix is shown in Fig.~\ref{static}(b) by comparing our results and the TLL prediction \cite{note_1}.
The TLL approach does not work because the temperature considered here is beyond its validity; see Part E of the Supplemental Material.

In conclusion, a detailed study on the finite temperature 1BDCF of the Lieb-Liniger model with variable repulsive interaction strengths has been presented by using the determinant expression of form factor and numerics.
In particular, we have considered the regime far away from the GP and TG limit \cite{Kheruntsyan}.
By comparing the situations of the 1BDCF at zero and finite temperatures, the results of spectral distribution, line shape, the momentum distribution, and one-body density matrix have demonstrated the subtle competition between thermal fluctuation and quantum interaction.
The thermal fluctuation destroys the sharp edge singularity behavior of single-particle spectrum, especially removing the singularity at the hole-induced edge, i.e., type-II dispersion relation.
It drives the distribution of spectral weight spread in the negative energy plane, clearly reflected by the line shape of 1BDCF.
The double-peaks structure of 1BDCF can be explained by a series of "relative excitation" over the reference state, as well as by the perspective of the detailed balancing relation.
Our results of the one-body density matrix manifest the intensity of thermal fluctuation in the region considered strong enough to break down the validity of TLL, demonstrating a broadly applicable range of our method.
It is worth emphasizing that our method  can be further generalized into the realistic experimental setting --- the gases of cold atoms are trapped in a harmonic potential.
This problem can be treated as a uniform case together with local density approximation.
The capability of our method for dealing with arbitrary interaction strength and a wide range of temperatures promises a reliable benchmark for the cutting-edge cold atom experiments. Moreover, our method can be employed in the study of dynamical fermionization \cite{Rigol2005,Alam2021,Wilson2020,Patu2023} and interaction quench.

{\em Acknowledgments.}
This work is supported by the National Natural Science Foundation of China Grants No. 12104372, No. 12088101, No. 12047511, No. 12247103, No. 12134015, and No. 92365202, HK GRF Grants No. 17306024 and No. 17313122, CRF Grants No. C4050-23GF and No. C7012-
21G, and RGC Fellowship Award No. HKU RFS2223-7S03. 
S.C. and Y.Y.C. contributed equally to this work.

{\em Note Added.}
After the release of this work on arXiv: 2211.00282, we became aware of recently another paper \cite{Caux2023} that discusses the improvements on ABACUS \cite{Caux2009} and the difficulty in evaluating 1BDCF at finite temperatures.

\begin{figure*}[ht]
 \centering
 \includegraphics[width=\textwidth]{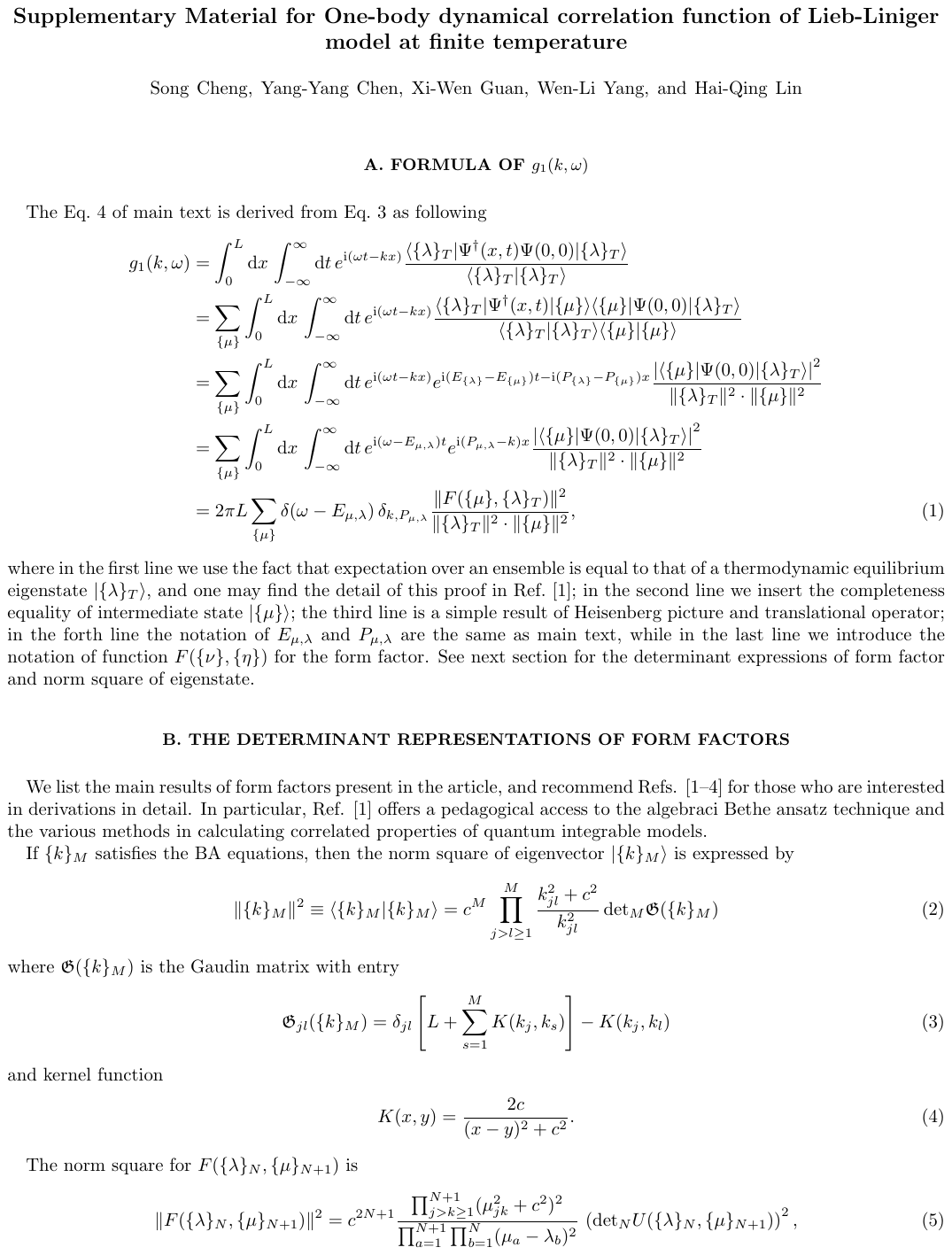}
\end{figure*}

\begin{figure*}[ht]
 \centering
 \includegraphics[width=\textwidth]{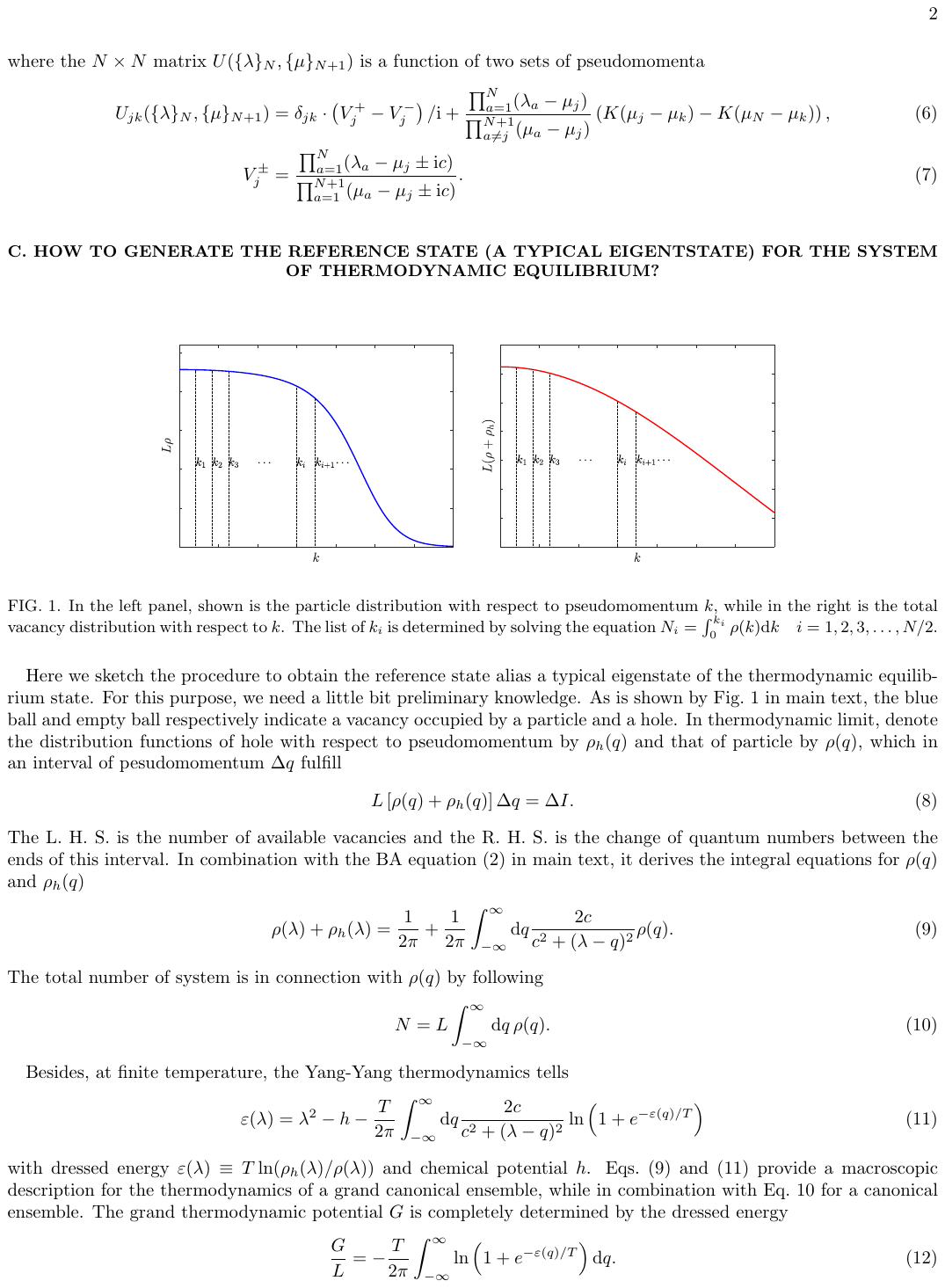}
\end{figure*}

\begin{figure*}[ht]
 \centering
 \includegraphics[width=\textwidth]{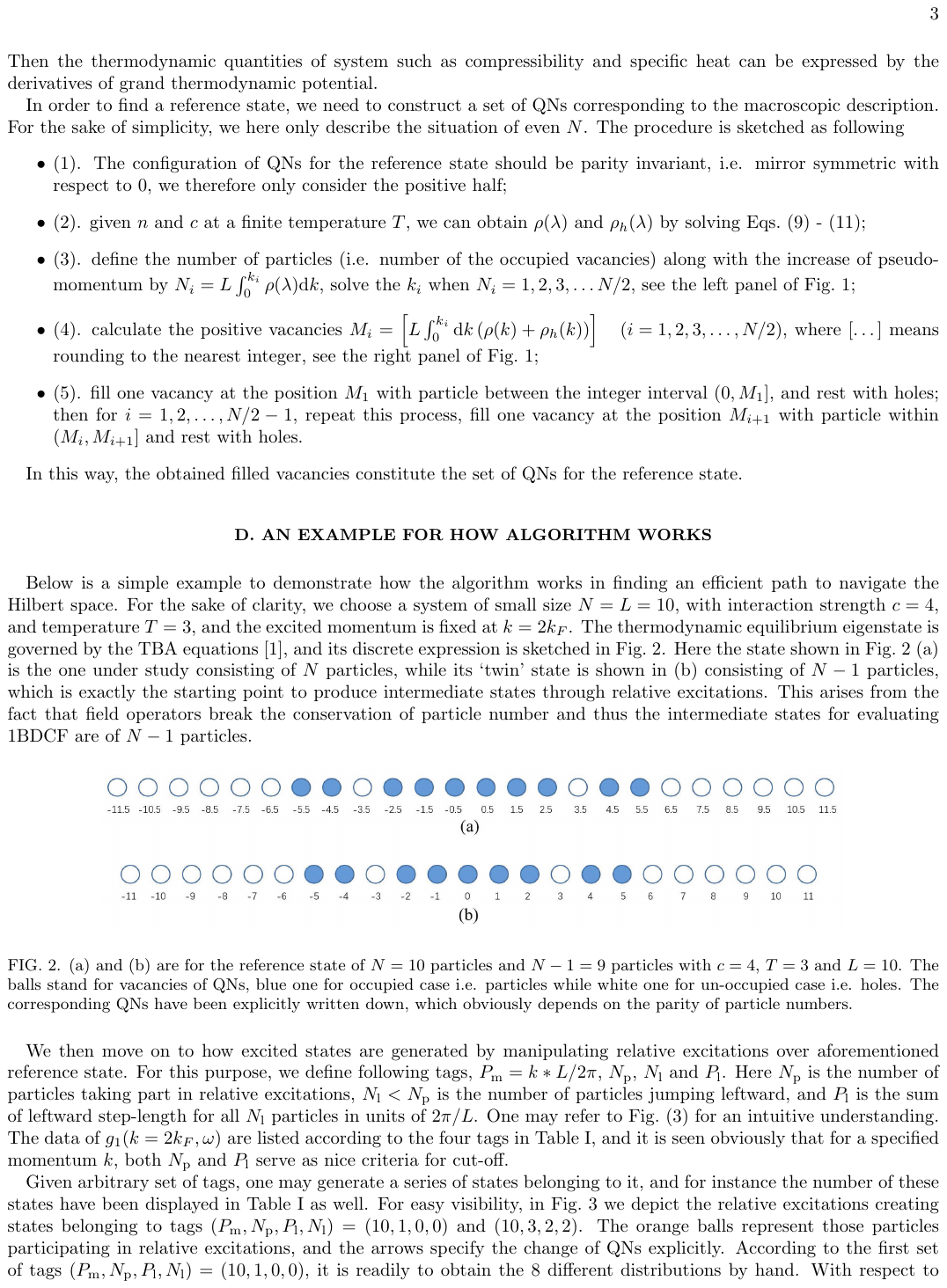}
\end{figure*}

\begin{figure*}[ht]
 \centering
 \includegraphics[width=\textwidth]{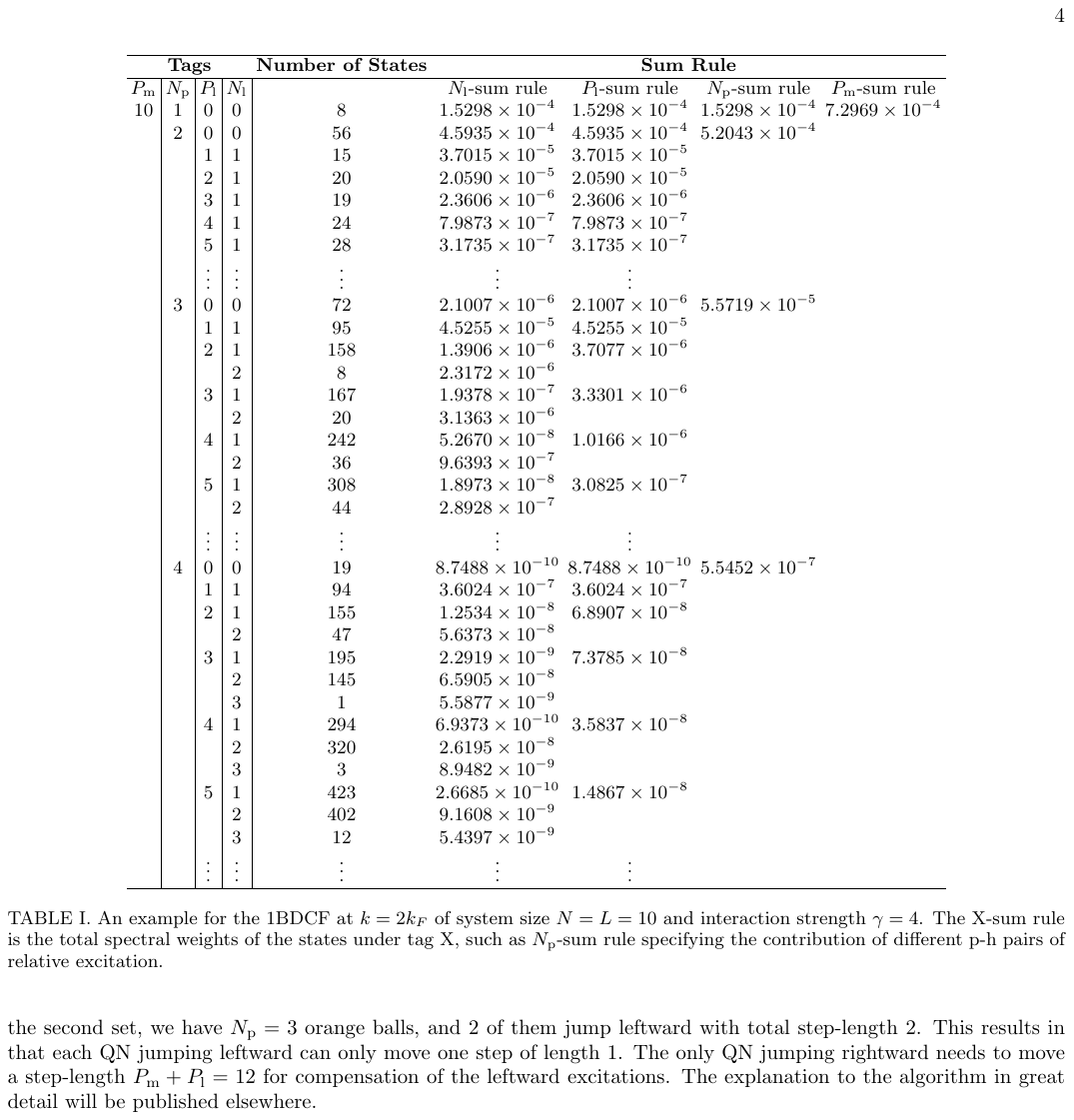}
\end{figure*}

\begin{figure*}[ht]
 \centering
 \includegraphics[width=\textwidth]{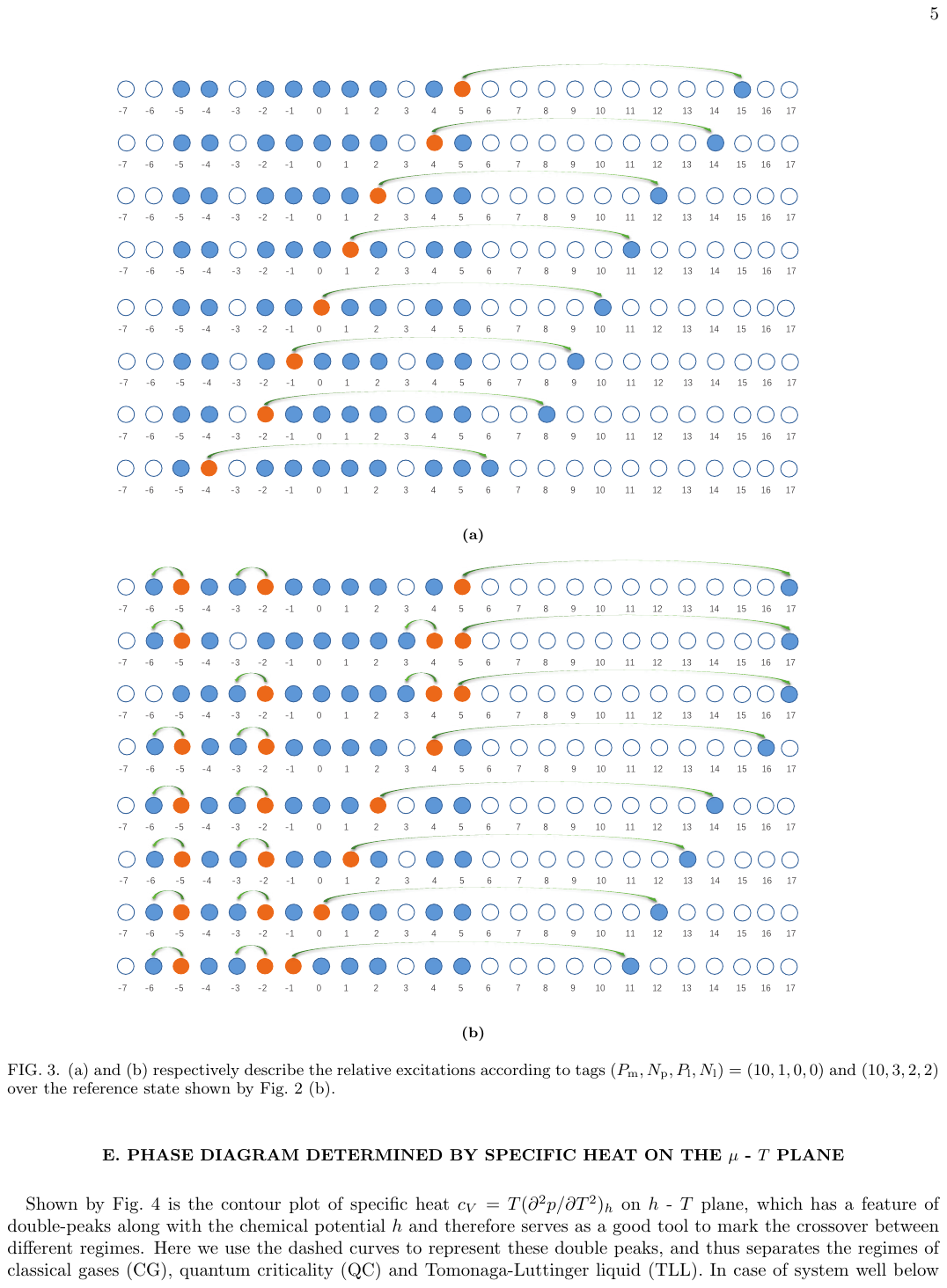}
\end{figure*}

\begin{figure*}[ht]
 \centering
 \includegraphics[width=\textwidth]{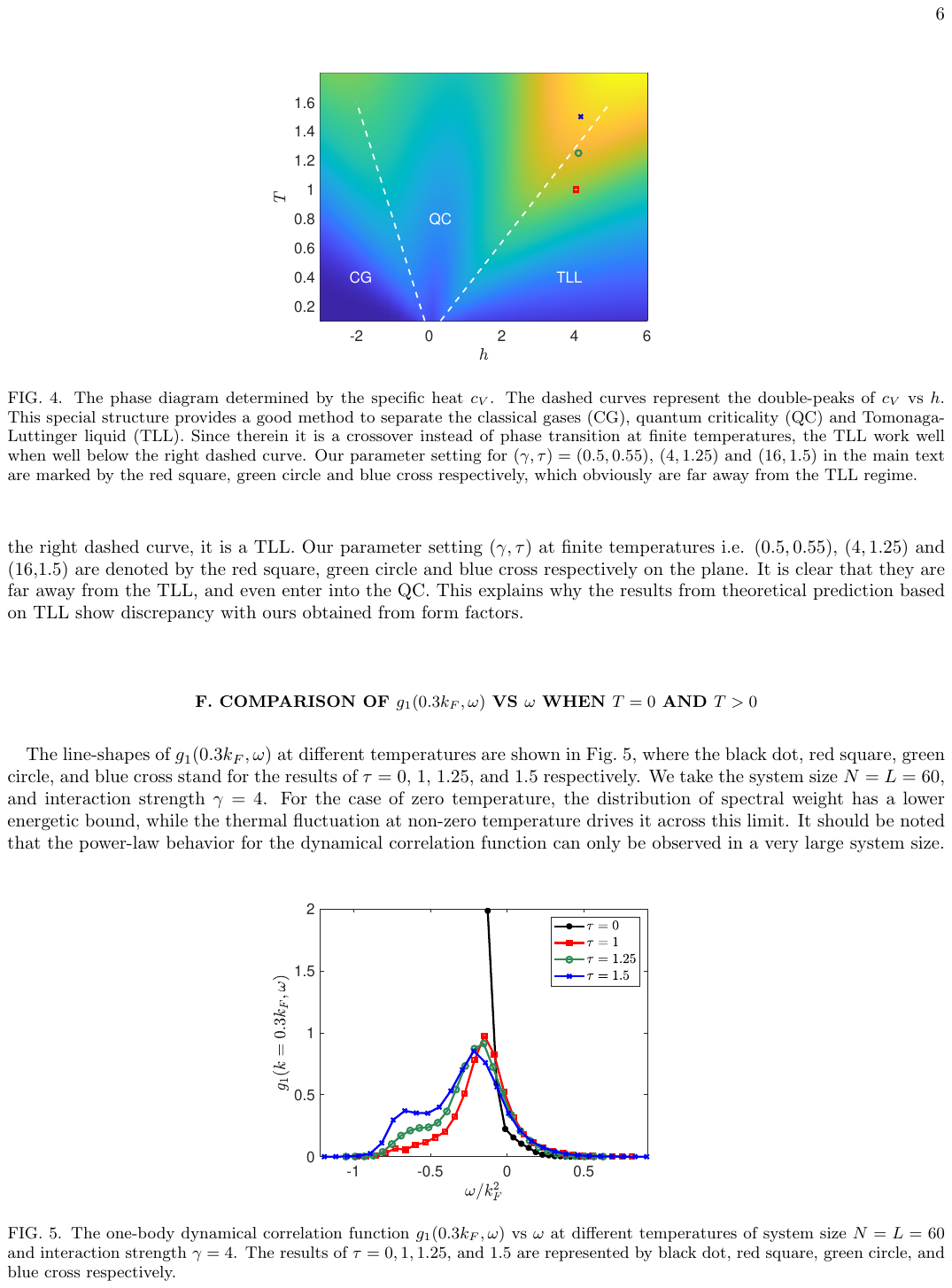}
\end{figure*}

\begin{figure*}[ht]
 \centering
 \includegraphics[width=\textwidth]{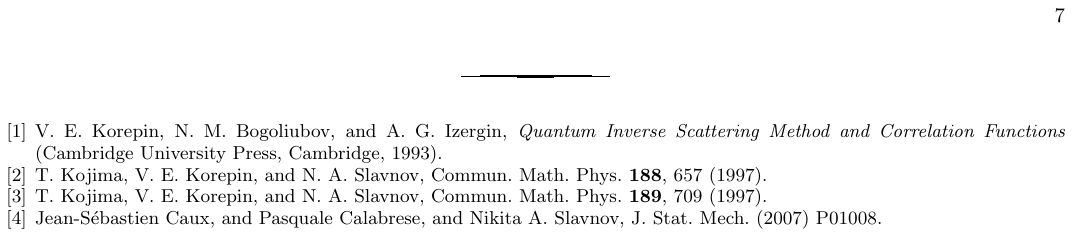}
\end{figure*}


\begin{thebibliography}{99}

\bibitem{Cazalilla}
M. A. Cazalilla, R. Citro, T. Giamarchi, E. Orignac, and M. Rigol, One dimensional bosons: From condensed matter systems to ultracold gases, 
Rev. Mod. Phys. {\bf 83}, 1405 (2011).

\bibitem{Cazalilla2004}
M. A. Cazalilla, Bosonizing one-dimensional cold atomic gases, J. Phys. B. {\bf 37}, S1 (2004).

\bibitem{Imambekov2012}
A. Imambekov, T. L. Schmidt and L. I. Glazman, One-dimensional quantum liquids: Beyond the Luttinger liquid paradigm, Rev. Mod. Phys. {\bf 84}, 1253 (2012).

\bibitem{Guan2013}
X.-W. Guan, M. T. Batchelor, and C.-H. Lee, Fermi gases in one dimension: From Bethe ansatz to experiments, Rev. Mod. Phys. {\bf 85}, 1633 (2013).

\bibitem{Guan2022}
X.-W. Guan and P. He, New trends in quantum integrability: recent
experiments with ultracold atoms, Rep. Prog. Phys. {\bf 85}, 114001 (2022).

\bibitem{Landau:1957} L. D. Landau, On the theory of the Fermi liquid, JETP {\bf 3}, 920 (1957).

\bibitem{Anderson:1984} P. W. Anderson, {\em Basic Notions of Condensed Matter Physics} (Benjamin Cummings, Penlo Park, 1984).

\bibitem{Haldane:1981} F. D. M. Haldane, Effective Harmonic-Fluid Approach to Low-Energy Properties of One-Dimensional Quantum Fluids, Phys. Rev. Lett. {\bf 47}, 1840 (1981); `Luttinger liquid theory' of one-dimensional quantum
fluids: I. Properties of the Luttinger model and their extension to the general 1D interacting spinless Fermi gas, J. Phys. C {\bf 14}, 2585 (1981).

\bibitem{Giamarchi}
T. Giamarchi, {\em Quantum Physics in One  Dimension} (Oxford University Press, Oxford, 2004).


\bibitem{Giorgini:2008} S. Giorgini, L. P.  Pitaevskii and S. Stringari, Theory of ultracold atomic Fermi gases, Rev. Mod. Phys. {\bf 80}, 1215 (2008).

\bibitem{Ess05} F. H. L. Essler, H. Frahm, F. G\"{o}hmann, A. Kl\"{u}mper  and V. E. Korepin,
\textit{The One-Dimensional Hubbard Model} (Cambridge University Press, Cambridge, 2005).

\bibitem{Hilker:2017}T. A. Hilker, G. Salomon,  F. Grusdt, A. Omran, M. Boll, E.  Demler, I. Bloch, C. Gross, Revealing hidden antiferromagnetic correlations in doped Hubbard chains via string correlators, Science {\bf 357}, 484 (2017).

\bibitem{Vijayan:2020} J. Vijayan, P. Sompet, G. Salomon, J. Koepsell, S. Hirthe, A. Bohrdt, F. Grusdt, I. Bloch, and C. Gross, Time-resolved observation of spin-charge deconfinement in fermionic Hubbard chains, Science {\bf 367}, 186 (2020).

\bibitem{Senaratne:2022} R. Senaratne, D. Cavazos-Cavazos, S. Wang, F. He, Y.-T. Chang, A. Kafle, H. Pu, X.-W. Guan, and R. G. Hulet, Spin-charge separation in a one-dimensional Fermi gas with tunable interactions, Science {\bf 376}, 1305 (2022).

\bibitem{Kinoshita:2006}T. Kinoshita, T. Wenger, and D. S. Weiss, A quantum Newton's cradle, Nature {\bf 440}, 900 (2006).

\bibitem{Langen:2015}T. Langen, S. Erne, R. Geiger, B. Rauer, T. Schweigler, M. Kuhnert, W. Rohringer, I. E. Mazets, T. Gasenzer, and J. Schmiedmayer, Experimental Observation of a Generalized Gibbs Ensemble, Science {\bf 348}, 207 (2015).

\bibitem{Wilson2020}
Joshua M. Wilson, Neel Malvania, Yuan Le, Yicheng Zhang, Marcos Rigol, David S. Weiss, Observation of dynamical fermionization, Science {\bf 367}, 1461 (2020).


\bibitem{Lieb}
E. H. Lieb, and W. Liniger, Exact Analysis of an Interacting Bose Gas. I. The General Solution and the Ground State,Phys. Rev. {\bf 130}, 1605 (1963); E. H. Lieb, Exact Analysis of an Interacting Bose Gas. II. The Excitation Spectrum, \textit{ibid.} {\bf  130}, 1616  (1963).

\bibitem{Kheruntsyan}
K. V. Kheruntsyan, D. M. Gangardt, P. D. Drummond, and G. V. Shlyapnikov, Pair Correlations in a Finite-Temperature 1D Bose Gas, Phys. Rev. Lett. {\bf 91}, 040403 (2003); Finite-temperature correlations and density profiles of an inhomogeneous interacting one-dimensional Bose gas, Phys. Rev. A {\bf 71}, 053615 (2005).

\bibitem{Guan:2011} X.-W. Guan and M. T. Batchelor, Polylogs, thermodynamics and scaling functions of one-dimensional quantum many-body systems, J. Phy. A {\bf 44}, 102001 (2011).

\bibitem{Girardeau}
M. Girardeau, Relationship between Systems of Impenetrable Bosons and Fermions in One Dimension, J. Math. Phys. {\bf 1}, 516 (1960); Permutation Symmetry of Many-Particle Wave Functions, Phys. Rev. {\bf 139}, B500 (1965).

\bibitem{Lenard}
A. Lenard, Momentum Distribution in the Ground State of the One-Dimensional System of Impenetrable Bosons, J. Math. Phys. {\bf 5}, 930 (1964); One Dimensional Impenetrable Bosons in Thermal Equilibrium, \textit{ibid.} {\bf 7}, 1268 (1966).

\bibitem{Tracy}
H. G. Vaidya, and C. A. Tracy, One Particle Reduced Density Matrix of Impenetrable Bosons in One dimension at Zero Temperature, Phys. Rev. Lett. {\bf 42}, 3 (1979); One particle reduced density matrix of impenetrable bosons in one dimension at zero temperature, J. Math. Phys. {\bf 20}, 2291 (1979).

\bibitem{Pezer}
R. Pezer and H. Buljan, Momentum Distribution Dynamics of a Tonks-Girardeau Gas: Bragg Reflections of a Quantum Many-Body Wave Packet, Phys. Rev. Lett. {\bf 98}, 240403 (2007).

\bibitem{Minguzzi}
J. Settino, N. Lo Gullo, F. Plastina, and A. Minguzzi, Exact Spectral Function of a Tonks-Girardeau Gas in a Lattice, Phys. Rev. Lett. {\bf 126}, 065301 (2021).


\bibitem{Colcelli2018a}
A. Colcelli, J. Viti,  G. Mussardo, and A. Trombettoni, Universal off-diagonal long-range-order behavior for a trapped Tonks-Girardeau gas, Phys. Rev. A {\bf98}, 063633 (2018).

\bibitem{Colcelli2018b}
A. Colcelli,  G. Mussardo, and A. Trombettoni, Deviations from off-diagonal long-range order in one-dimensional quantum systems, Euro. Phys. Lett. {\bf122}, 50006 (2018).

\bibitem{Colcelli2020}
A. Colcelli, N. Defenu, G. Mussardo, and A. Trombettoni, Finite temperature off-diagonal long-range order for interacting bosons, Phys. Rev. B {\bf102}, 184510 (2020).


\bibitem{Cherny}
J. Brand, and A. Yu. Cherny, Dynamical Structure Factor of the One-Dimensional Bose Gas Near the Tonks-Giradeau Limit, Phys. Rev. A {\bf 72}, 033619 (2005); A. Yu. Cherny and J. Brand, Polarizability and dynamic structure factor of the one-dimensional Bose gas near the Tonks-Girardeau limit at finite temperature, \textit{ibid.} {\bf 73}, 023612 (2006).

\bibitem{Gora_a}
D. M. Gangardt and G. V. Shlyapnikov, Stability and Phase Coherence of Trapped 1D Bose Gases, Phys. Rev. Lett. {\bf 90}, 010401 (2003).

\bibitem{Gora_b}
D. M. Gangardt and G. V. Shlyapnikov, Local correlations in a strongly interacting one-dimensional Bose gases, New. J. Phys. {\bf  5}, 79 (2003).

\bibitem{Nandani}
E. J. K. P. Nandani, R. A. R\"{o}mer, S.-N. Tan, and X.-W. Guan, Higher-Order Local and Non-Local Correlations for 1D Strongly Interacting Bose Gas, New. J. Phys. {\bf 18}, 055014 (2016).

\bibitem{Xu2015}
W. Xu and M. Rigol, Universal scaling of density and momentum distributions in Lieb-Liniger gases, Phys. Rev. A {\bf 92}, 063623 (2015).

\bibitem{Rosi2023}
G. De Rosi, R. Rota, G. E. Astrakharchik, and J. Boronat, Correlation properties of a one-dimensional repulsive Bose gas at finite temperature, New. J. Phys. {\bf 25}, 043002 (2023).


\bibitem{Imambekov2008}
A. Imambekov, and L. I. Glazman, Exact Exponents of Edge Singularities in Dynamic Correlation Functions of 1D Bose Gas, Phys. Rev. Lett. {\bf 100}, 206805 (2008); Phenomenology of One-Dimensional Quantum Liquids Beyond the Low-Energy Limit, \textit{ibid.} {\bf 102}, 126405 (2009).

\bibitem{Franchinni}
F. Franchini, {\em An Introduction to Integrable Techniques for One-Dimensional Quantum Systems} (Springer,  Cham, 2017)

\bibitem{QISM}
V. E. Korepin,  N. M. Bogoliubov, and A. G. Izergin, \textit{Quantum Inverse Scattering Method and Correlation Functions}  (Cambridge University Press, Cambridge, 1993).

\bibitem{Korepin1982}
V. E. Korepin, Calculation of Norms of Bethe Wave Functions, Commun. Math. Phys. {\bf 86}, 391 (1982).

\bibitem{Korepin1984}
V. E. Korepin, Correlation functions of the one-dimensional Bose gas in the repulsive case, Commun. Math. Phys. {\bf 94}, 93 (1984).

\bibitem{Slavnov}
N. A. Slavnov, Calculation of scalar products of wave functions and form factors in the framework of the algebraic Bethe ansatz, Theor. Math. Phys. {\bf 79}, 502 (1989); Nonequal-time current correlation function in a one-dimensional Bose gas, \textit{ibid.} {\bf 82}, 273 (1990).

\bibitem{Slavnov1997}
T. Kojima, V. E. Korepin, and N. A. Slavnov, Determinant Representation for Dynamical Correlation Functions of the Quantum Nonlinear Schrodinger Equation, Commun. Math. Phys. {\bf 188}, 657 (1997); Completely Integrable Equation for Quantum Correlation Functions of Nonlinear Schrodinger Equation, {\em ibid.} {\bf 189}, 709 (1997).

\bibitem{Caux2006}
J.-S. Caux, and P. Calabrese, Dynamical Density-Density Correlations in the One-Dimensional Bose Gas, Phys. Rev. A {\bf 74}, 031605(R) (2006).

\bibitem{Caux2007}
J.-S. Caux, P. Calabrese, and N. A. Slavnov, One-particle dynamical correlations in the one-dimensional Bose gas, J. Stat. Mech. (2007) P01008.

\bibitem{Panfil}
M. Panfil, and J.-S. Caux, Finite-temperature correlations in the Lieb-Liniger one-dimensional Bose gas, Phys. Rev. A {\bf 89}, 033605 (2014).

\bibitem{Caux2009}
J.-S. Caux, Correlation functions of integrable models: a description of the ABACUS algorithm, J. Math. Phys. {\bf 50}, 095214 (2009).

\bibitem{SC}
S. Cheng, Y.-Y. Chen, X.-W. Guan, W.-L. Yang, R. Mondaini, and H.-Q. Lin, Exact Spectral Function of One-Dimensional Bose Gases, arXiv:2209.15221v1.

\bibitem{Caux2023}
A. J. J. M. de Klerk, and J.-S. Caux, Improved Hilbert space exploration algorithms for finite temperature calculations, SciPost. Phys. Core {\bf 6}, 039 (2023); arXiv: 2301.09224.

\bibitem{YYC2023}
R.-T. Li, S. Cheng, Y.-Y. Chen and X.-W. Guan, Exact results of dynamical structure factor of Lieb–Liniger model, J. Phys. A. {\bf 56}, 335204 (2023).


\bibitem{Exp}
N. Fabbri, D. Cl\'{e}ment, L. Fallani, C. Fort and M. Inguscio, Momentum-resolved study of an array of one-dimensional strongly phase-fluctuating Bose gases, Phys. Rev. A {\bf 83}, 031604(R) (2011);
F. Meinert, M. Panfil, M. J. Mark, K. Lauber, J.-S. Caux, and H.-C. N\"{a}gerl, Probing the Excitations of a Lieb-Liniger Gas from Weak to Strong Coupling, Phys. Rev. Lett. {\bf 115}, 085301 (2015);
B. Yang, Y.-Y. Chen, Y.-G. Zheng,  H. Sun, H.-N. Dai, X.-W. Guan, Z.-S. Yuan, and  J.-W. Pan, Quantum criticality and the Tomonaga-Luttinger liquid in one-dimensional Bose gases, Phys. Rev. Lett. {\bf 119}, 165701 (2017); 


\bibitem{Granet}
E. Granet and F. H. L. Essler, A systematic $1/c$-expansion of form factor sums
for dynamical correlations in the Lieb-Liniger model, SciPost. Phys. {\bf 9}, 082 (2020);
E. Granet, Low-density limit of dynamical correlations in the Lieb–Liniger model, J. Phys. A. {\bf 54}, 154001 (2021).

\bibitem{Shashi}
A. Shashi, L. I. Glazman, J.-S. Caux, and A. Imambekov, Nonuniversal prefactors in the correlation functions of one-dimensional quantum liquids, Phys. Rev. B {\bf 84}, 045408 (2011);
A. Shashi, M. Panfil, J.-S. Caux, and A. Imambekov, Exact prefactors in static and dynamic correlation functions of one-dimensional quantum integrable models: Applications to the Calogero-Sutherland, Lieb-Liniger, and XXZ models, Phys. Rev. B {\bf 85}, 155136 (2012).


\bibitem{YY}
C. N. Yang and C. P. Yang, Thermodynamics of a One‐Dimensional System of Bosons with Repulsive Delta‐Function Interaction, J. Math. Phys. {\bf 10}, 1115 (1969).

\bibitem{note4SM}
See Supplemental Material for the determinant expression of form factor, how to generate a reference state, example of algorithm, and phase diagram at finite temperature, which
includes Refs. [37,41,43].

\bibitem{Zambelli2000}
F. Zambelli, L. Pitaevskii, D. M. Stamper-Kurn, and S. Stringari, Dynamic structure factor and momentum distribution of a trapped Bose gas, Phys. Rev. A {\bf 61}, 063608 (2000).


\bibitem{note}
See Eq.~(I.4.29) of  Ref.~\cite{QISM} for more explanation.

\bibitem{LL}
L. D. Landau and E. M. Lifshitz, \textit{Statistical Physics} Part 2, (Pergamon Press, Oxford, UK 1981).

\bibitem{note_1}
The TLL prediction comes from following formula cf. Ref.~\cite{Cazalilla2004},
\begin{align}\label{TLL_rho_k}
n (k,T) \approx & \, \left( \frac{2K}{\pi} \right)^{\frac{1}{2K}} \left( \frac{n L_\phi}{K} \right)^{1-\frac{1}{2K}} \Gamma\left(1-\frac{1}{2K}\right)  \notag\\
&\cdot \textmd{Re}\left[ \frac{ \Gamma( \frac{1}{4K} + \textmd{i} \frac{k L_\phi}{2K}) } { \Gamma(1-\frac{1}{4K} + \textmd{i} \frac{k L_\phi}{2K}) } \right] \nonumber
\end{align}
where $m$ is the particle mass, $K$ the Luttinger parameter, and  $L_\phi = \hbar^2 n/(m k_B T)$ is the phase correlation length.

\bibitem{Rigol2005}
M. Rigol and A. Muramatsu, Fermionization in an Expanding 1D Gas of Hard-Core Bosons, Phys. Rev. Lett. {\bf 94}, 240403
(2005).

\bibitem{Alam2021}
Shah Saad Alam, Timothy Skaras, Li Yang, and Han Pu, Dynamical Fermionization in One-Dimensional Spinor Quantum Gases, Phys. Rev. Lett. {\bf 127}, 023002 (2021).

\bibitem{Patu2023}
Ovidiu I. Patu, Dynamical Fermionization in One-Dimensional Spinor Gases at Finite Temperature, Phys. Rev. Lett. {\bf 130}, 163201 (2023).

\end{thebibliography}
\end{document}